\documentstyle[12pt,epsf]{article}
\newcommand{\bc}{\begin{center}}
\newcommand{\ec}{\end{center}}
\newcommand{\be}{\begin{equation}}
\newcommand{\ee}{\end{equation}}
\newcommand{\ba}{\begin{array}}
\newcommand{\ea}{\end{array}}
\newcommand{\bea}{\begin{eqnarray}}
\newcommand{\eea}{\end{eqnarray}}
\newcommand{\bt}{\begin{tabular}}
\newcommand{\et}{\end{tabular}}

\begin{document}

\vspace{1cm}

\begin{center}
{\Large\bf
The bound state corrections \\[1mm]
to the semileptonic decays of the $B$ meson. \\[2mm]
The light--front approach versus ACM model.}
\\[1cm]
{\bf I.L.Grach, P.Yu.Kulikov and I.M.Narodetskii}\\[0.5cm]
Institute of the Theoretical and Experimental Physics, Moscow\\
\end{center}
\vspace{2cm}

\begin{abstract}
\noindent A generalization of the parton--like formula is used for the first
time to find the
differential distributions in the inclusive semileptonic weak decays of the
$B$ meson. The main features of this new approach are the treatment of
the $b$--quark as an on--mass--shell particle and the inclusion of the
effects arising from the $b$--quark transverse motion in the $B$--meson.
Using the $b$--quark light--front (LF) distribution function related to
the equal time momentum wave function  taken from the ACM model
we compute the electron energy spectra and the total semileptonic
widths of the $B$ meson.
We find an impressive agreement between the electron energy
spectra calculated in  the LF approach and the ones obtained in the
ACM model, provided the $b$ quark mass is identified with the average of the
floating $b$ quark mass in the ACM model. In spite of the simplicity of
the model we obtain a fair good description of the CLEO data for
$|V_{cb}|=0.042$.
\end{abstract}

\vspace{3cm}


\vspace{0.5cm}


\newpage
\noindent{\it 1. Introduction}. Studying of the leptonic  spectra
in the weak semileptonic $B\to X_c(X_u)\ell\nu_{\ell}$
decays\footnote{Throughout this paper we use the charge conjugated
notations.} is important to extract $|V_{cb}|$ and $|V_{ub}|$. The
latter plays an important role in the determination of the
unitarity triangle. Since the $b$--quark is heavy compared to the
QCD scale, the inclusive semileptonic $B$ decays can be treated
with the help of an operator product expansion (OPE) combined with
the heavy quark expansion (HQE) \cite{BSU97}. The result (away
from the endpoint of the spectrum) is that the inclusive
differential decay  width $d\Gamma/dE$ may be  expanded in
$\Lambda/m_b$, where $\Lambda$ is a QCD related scale of order 500
MeV, and $m_b$ is the mass of the heavy quark. The leading term
(zeroth order in $\Lambda/m_b$) is the free quark decay spectrum,
the subleading term vanishes, and the subsubleading term involves
parameters from the heavy quark theory, but should be rather
small, as it is  of order $(\Lambda/m_b)^2$. However, near the end
point the $1/m_b$ expansion has to be replaced by an expansion in
twist. To describe this region one has to introduce a so--called
`shape function', which in principle introduces a large hadronic
uncertainty. This is quite analogous to what happens for the
structure function in deep--inelastic scattering in the region
where the Bjorken variable $x_B\to 1$. A model independent
determination of the shape function is not available at the
present time, therefore a certain model dependence in this region
seems to be unavoidable, unless lattice data become reasonably
precise.

As to phenomenological analyzes of the photon spectra up to now
they have been solely based on the ACM model \cite{AP79},
\cite{ACM82}.
The various light--front (LF) approaches to consideration of the inclusive semileptonic
transitions were suggested in Refs. \cite{JP94}--\cite{KNST99}.
In Refs. \cite{JP94}, \cite{MTM96} the Infinite Momentum Frame
prescription  $p_b=\xi P_B$, and, correspondingly, the floating $b$ quark mass
$m_b^2(\xi)=\xi^2M_B^2$ have been used. The transverse $b$ quark momenta
were consequently neglected.
In Ref. \cite{KNST99} the $b$--quark was considered as an on--mass--shell
particle with the definite mass $m_b$ and
the effects arising from the $b$--quark transverse motion in the
$\bar B$--meson were included. The corresponding ans\"atz of Ref. \cite{KNST99} reduces to a specific
choice of
the primordial LF distribution function $|\psi(\xi,p^2_{\bot})|^2$,
which represents the probability to find the
$b$ quark carrying a LF
fraction $\xi$ and a transverse momentum squared $p^2_{\bot}=|{\bf p}_{\bot}|^2$.
As a result, a new parton--like formula for the inclusive semileptonic
$b\to c,u$ width has been derived \cite{KNST99}, which is similar
to the one obtained by Bjorken {\it et al.} \cite{BDT92} in case of infinitely heavy
$b$ and $c$ quarks.

In this paper, we use the techniques developed in Ref.
\cite{KNST99} to evaluate the non--perturbative corrections to the
lepton spectrum in the inclusive $B\to X_c\ell\nu_{\ell}$
decays. We strive to implement the binding and the $B$--meson wave
function effects on the lepton energy spectrum. The main purpose
of the paper is to confront the lepton spectra and the total
semileptonic decay widths calculated in the LF and ACM approaches.
We find that the discrepancy between the two is very small
numerically.\\[2mm]

\noindent{\it 2. ACM versus LF}.
The decay spectrum in the ACM model is determined by the
kinematics constrains on the $b$
 quark. It incorporates some of the corrections related to the fact that the decaying $b$ quark
is not free, but in a bound state. The model treats the $B$ meson with the mass $m_B$ as
consisting of the
heavy $b$ quark plus a spectator with fixed mass $m_{sp}$; the latter usually represents a
fit parameter. The spectator quark has a momentum distribution
$\phi (|{\bf p}|)$ where ${\bf p}$ is its three-dimensional
momentum.  The momentum distribution is usually taken to be  Gaussian:
\be
\label{1}
\phi(|{\bf p}|)={4 \over \sqrt{\pi} p_f^3}\exp\left(-{|{\bf p}|^2\over p_f^2}\right),
\ee
which is normalized so that the integral over all momenta
of $\phi(|{\bf p}|) p^2$  is 1.
The energy-momentum conservation in the $B$ meson vertex implies
that the $b$ quark energy $E_b=m_B-\sqrt{{\bf p}^2+m_{sp}^2}$, where
$m_B$ is the mass of the $B$ meson;
thus the $b$ quark cannot possess a definite mass. Instead, one obtains
a ``floating'' $b$ quark mass
$(m_b^f)^2=m_B^2+m_{sp}^2-2m_B\sqrt{{\bf p}^2+m_{sp}^2}$~,
which depends on $|{\bf p}|$. The lepton spectrum is first obtained from
the spectrum $d\Gamma_b^{(0)}(m_f,E')/dE'$ of the $b$ quark
of invariant mass $m_b^f$ (in the $b$ quark rest frame) and then
boosting back to the rest frame of the $B$ meson
and averaging over the weight function $\phi (|{\bf p}|)$ (further details can be found in
the original work \cite{ACM82}).


The LF approach differs from that of the ACM model in two respects. First,
similar to the ACM model the LF quark
model treats the beauty meson $B$ as consisting of the heavy $b$ quark plus a
spectator quark. Both quarks have fixed masses, $m_b$ and $m_{sp}$, though.
This is at variance with the ACM model, that has been introduced in order
to avoid the notion of the heavy quark mass at all. Secondly,
the calculation of the distribution over lepton energy
in the LF approach does not requires  any boosting procedure but instead is
based on  the standard Lorentz--invariant kinematical analysis (see {\it e.g.}
Ref. \cite{BKSV93}), which we now briefly discuss.


There are three independent kinematical variables in the inclusive
phenomenology, for which we choose the lepton energy $E_{\ell}, q^2$,
where $q=p_{\ell}+p_{\nu_{\ell}}$, and the invariant mass $M^2_X=(p_B-q)^2$
of a hadronic state.
Introducing the dimensionless variables
$y=2E_{\ell}/m_B$, $t=q^2/m^2_B$, and $s=M^2_X/m^2_B$,
the differential decay rate for semileptonic $B$ decay can be written as
\bea
\label{13}
\frac{d\Gamma_{SL}}{dy}&=& \frac{G_F^2m_B^5}{64\pi^3}|V_{cb}|^2
\int\limits^{t_{
max}}_{0}dt\int\limits^{s_{max}}_{s_0}ds\nonumber \\
&&\times\left\{tW_1+\frac{1}{2}[y(1+t-s)-y^2-t]W_2+
t[\frac{1+t-s}{2}-y]W_3+\ldots\right\},
\eea
where the structure functions $W_i=W_i(s,t)$ appear in the decomposition
of the hadronic tensor $W_{\alpha\beta}$ in Lorentz covariants
\cite{BKSV93}.
The ellipsis in (\ref{13}) denote the terms proportional to the lepton mass
squared. The kinematical limits of integration can be found from equation
\be
\label{14}
\frac{s}{1-y}+\frac{t}{y}\leq 1.
\ee
They are given by $0\leq y\leq y_{max}=1-\left(m_{X_c}^{(min)}/m_B\right)^2$, where $m_{X_c}^{(min)}$ is the minimal mass
of the hadronic state $X_c$, $s_{max}=1+t-(y+t/y)$,
and $t_{max}=y\left( (1-(1-y_{max})/(1-y)\right)$.
\vspace{2mm}

\noindent {\it 3. Light--cone distribution functions}.
In a parton model we treat inclusive semileptonic $ B \to X_c\ell\nu_{\ell} $
decay in a direct analogy to deep-inelastic scattering. Specifically, we assume that the sum over all possible charm final
states $X_c$ can be modeled by the decay width of an on--shell $b$ quark
into on--shell $c$--quark weighted with the $b$--quark distribution.
Following the above assumption, the hadronic tensor $ W_{\alpha\beta} $ is written as
\be
\label{16}
W_{\alpha\beta}=\int w^{(cb)}_{\alpha\beta}(p_c,p_b)\delta
[(p_b-q)^2-m_c^2]\frac{|\psi(\xi,p^2_{\bot})|^2}{\xi}\theta(\varepsilon_c)
d\xi d^2p_{\bot},
\ee
where
\be
\label{17}
    w^{(cb)}_{\alpha\beta}(p_c,p_b)  = \frac{1}{2} \sum_{spins}
    \bar{u}_{c} O_{\alpha} u_b \cdot \bar{u}_b O^+_{\beta}u_{c},
\ee
with $O_{\alpha}= \gamma_{\alpha}(1-\gamma_5)$. The factor $1/\xi$ in Eq. (\ref{16}) comes
from the normalization of the
$B$ meson vertex \cite{DGNS97}.


\noindent Eq. (\ref{16}) amounts to averaging the perturbative decay distribution
over motion of heavy quark governed by
the distribution function
$f(x,p^2_{\bot})=|\psi(x,p^2_{\bot})|^2$. In this respect our approach is similar to the parton model in deep inelastic scattering, although it is not really a parton model
in its standard definition. The normalization condition reads
\be
\label{18} \pi\int\limits_0^1 d\xi\int
dp^2_{\bot}f(\xi,p^2_{\bot})=1. \ee The function
$\theta(\varepsilon_c)$ where $\varepsilon_c$ is the $c$--quark
energy is inserted in Eq. (\ref{16}) for consistency with the use
of valence LF wave function to calculate the $b$--quark
distribution in the $B$--meson.

Since we do not have an explicit representation for the B--meson Fock
expansion in QCD, we shall proceed by making an ansatz for
$\psi(\xi,p^2_{\bot})$. This is model dependent enterprise but has
its close equivalent in studies of electron spectra using the ACM
model. We  choose the momentum space structure
of an equal time (ET) wave function $\phi(|{\bf p}|)$ as in Eq. (\ref{1}).
We convert from ET to LF momenta by leaving the transverse
momenta unchanged and letting
\be
\label{19}
p_{iz}=\frac{1}{2}(p_i^+-p_i^-)=\frac{1}{2}(p_i^+-\frac{p^2_{i{\bot}}
+m^2_i}{p_i^+})
\ee
for both the $b$--quark $(i=b)$ and the quark--spectator $(i=sp)$. The longitudinal
LF momentum fractions $\xi_i$ are defined as $\xi_{sp}=p_{sp}^+/P_B^+$,
$\xi_b=p_b^+/P_B^+$, with $\xi_b+\xi_{sp}=1$. In the $B$--meson rest frame $P_B^+=m_B$.
Then for the distribution function $|\psi(\xi,p^2_{\bot})|^2~ (\xi=\xi_b)$ normalized
according to (\ref{18}) one obtains
\footnote{On the same footing one can consider $|{\bf p}|$ in (\ref{1}) as the
{\it relative} momentum between heavy and light quarks. In this case it is more
convenient to use the
quark--antiquark rest frame instead of the $B$--meson rest frame. Recall that
in the LF formalism these two frames are different. Then the longitudinal LF
momentum fractions $\xi_i$ are defined as $\xi_{sp}=p^+_{sp}/M_0$,
$\xi_b=p^+_b/M_0$, where the free mass $M_0$ is
$M_0=\sqrt{m^2_b+{\bf p}^2}+\sqrt{m^2_{sp}+{\bf p}^2}$
with  ${\bf p}^2=p^2_{\bot}+p^2_z$,~$p_z=(\xi-\frac{1}{2})M_0-(m^2_b-m^2_{sp})
/2M_0$.
In this case the explicit form of $|\partial p_z/\partial \xi|$ is given {\it e.g.}
by Eq. (10) of Ref. \cite{GNST97}. We have checked that {\it numerically} both
approaches yield identical results for the electron spectra}
\be
\label{21}
|\psi(\xi,p^2_{\bot})|^2= \frac{4}{\sqrt{\pi}p^3_f}\exp\left(-\frac{
p^2_{\bot}+p^2_z}{p^2_f}\right)\left|\frac{\partial p_z}{\partial\xi}\right|,
\ee
where
\be
\label{p_z}
p^2_z(\xi,p^2_{\bot})=\frac{1}{2}\left((1-\xi)m_B-\frac{p^2_{\bot}
+m^2_{{\rm sp}}}{(1-\xi)m_B}\right)~~~{\rm and}~~~
\left|\frac{\partial p_z}{\partial \xi}\right|=\frac{1}{2}\left(m_B+\frac{p^2_{\bot}
+m^2_{{\rm sp}}}{(1-\xi)^2m_B}\right).
\ee

The calculation of the structure functions $W_i(t,s)$ in the LF parton approximation
(\ref{16}) is straightforward. The result is
\be
\label{23}
W_i(t,s)=\int w_i(s,t,\xi,p^2_{\bot})\delta
[(p_b-q)^2-m_c^2]\frac{|\psi(\xi,p^2_{\bot})|^2}{\xi}\theta(\varepsilon_c)
d\xi d^2p_{\bot},
\ee
where $w_i(s,t,\xi,p^2_{\bot})$ are the structure functions for
 the free quark decay.
For further details see Appendix of \cite{KNST99}.
Eq. (\ref{23}) differs from the corresponding expressions of Refs. \cite{JP94} and \cite{MTM96}
by the non--trivial dependence on $p^2_{\bot}$ which enters
both $|\psi(\xi,p^2_{\bot})|^2$ and argument of the $\delta$--function.
For further details see
\cite{KNST99}\footnote{Note that the expressions for $w_i$ in Ref. \cite{KNST99} miss an extra factor of 2.}.
\vspace{2mm}

\noindent {\it 4. Results}.
Having specified the non--perturbative aspects of our calculations, we
proceed to present numerical results for the lepton spectrum in the decay
$B\to X_c e\nu_e$. Our main computation refers to the case $m_{sp}=0.15$ GeV,
$m_c=1.5$ GeV as chosen in Ref. \cite{RS93}.

The choice of $m_b$ in our approach deserves some comments. In the ACM model,
it was shown \cite{RS93}, \cite{BSUV94} (see also \cite{LK96}) that the corrections to
first order in $1/m_b$ both
to the inclusive semileptonic width and to the regular part of the
lepton spectrum can be absorbed into the definition of the
quark mass: $m_b^{ACM}=<m_b^f>$, where $<m_b^f>$ is the value of the floating mass,
averaged over the distribution $\phi(|{\bf p}|)$.

The choice of $m_b$ in the LF approach was first addressed in the
context of the LF model for $b\to s\gamma$ transitions in Ref.
\cite{KKNS99}. Using the scaling feature of the photon spectrum in
the LF model, it was suggested that $m_b^{LF}$ can be defined from
the requirement of the vanishing of the first moment of the
distribution function. This condition coincides with that used in
HQE to define the pole mass of the $b$--quark. In this way one
avoids an otherwise large (and model dependent) correction of
order $1/m_b$ but at expense of introducing the shift in the
constituent quark mass which largely compensates the bound state
effects. It has been also demonstrated that the values of
$m_b^{LF}$ found by this procedure agree well with the average
values $<m_b^f>$ in the ACM model. The photon energy spectra calculated in
the LF approach were found to agree well with the ones obtained in the ACM model.

Accepting the identification $m_b^{LF}=m_b^{ACM}$ we want now to check
whether this result holds
for the description of the other channels, like $b\to c$. We  find again
a good agreement between the LF and ACM results but now for the semileptonic
$b\to c$ decays. The results of our computations of electron spectra and
the semileptonic widths are reported in Fig.1 and Table 1 using $|V_{cb}|=0.04$.
The different curves in Fig.1 correspond to the different values of $p_f$.
For each case we show separately the inclusive differential semileptonic
decay widths for the LF and ACM models and the free quark decay. The quark
decay spectra vanish for
$y>(m_b/m_B)(1-m_c^2/m_b^2)$, whereas the physical endpoint is $y_{max}=1-m^2_D/m_B^2$,
where  $m_D$ is the mass of the $D$ meson. In the LF  approach
the endpoint for the
electron spectrum is in fact not $y_{max}$ but
$y_{max}^{LF}=1-m^2_c/m_B^2$.
This is the direct consequence of the $p^2_{\bot}$ integration in
Eq. (\ref{23}) \cite{KNST99}. Note that $y_{max}^{LF}$ coincides with
$y_{max}^{ACM}$ with accuracy $\sim m_{sp}/m_B$.
For $m_c\sim 1.5$ GeV the difference between $y_{max}^{LF}\sim y_{max}^{ACM}$
and $y_{max}$ is of the order $10^{-2}$.
Another possibility advocated in \cite{GNST97} is to sum the electron spectra
from the exclusive $B\to D,D^*$ channels and from the
inclusive $B\to X_c$ channels, where $X_c$ is the hadronic
state with the mass $m_{X_c}\geq m_{D^*}$. Such the `hybrid' approach will
be considered elsewhere.

In Table 1 for various values of $p_f$, we give the corresponding
values of the total semileptonic width for the free quark with the
mass $m_b=<m_b^f>$ and the  $B$ meson semileptonic widths,
calculated using the LF and ACM approaches, respectively. In the
last two columns, we give the fractional deviation $\delta=\Delta
\Gamma_{SL}/\Gamma_{SL}^b$ (in per cent) between the semileptonic
widths determined in the LF and ACM models and that of the free
quark. The agreement between the LF and ACM approaches  for the
electron spectra is excellent for small $p_f$  as is exhibited in
Fig.1. A similar agreement also holds for integrated rates shown
in Table 1. This agreement is seen to be breaking down at $p_f\geq
0.4$ GeV, but even for $p_f\sim 0.5$ GeV the difference between
the ACM and LF inclusive widths is still small and is of the order
of a per cent level.

Finally, we calculate the $b\to c$ spectrum and compare it
with the experimental data from the CLEO
collaboration \cite{CLEO}. This is a direct calculation of the spectrum and not a $\chi^2$ fit.
We briefly investigated the sensitivity of the electron spectra to other
parameters of the models and found that the choice $p_f=0.4$ GeV, $m_c=1.5$ GeV $m_{sp}=0.15$
GeV is quite acceptable.

We display the results in Fig.2, where the three theoretical
curves are presented for the LF, ACM and free quark models. In
these calculations we have implicitly included the ${\it
O}(\alpha_s)$ perturbative corrections arising from gluon
Bremsstrahllung and one--loop effects which modify an electron
energy spectra at the partonic level (see {\it e.g.} \cite{JK89}
and references therein). It is customary to define a correction
function $G(x)$ to the electron spectrum $d\Gamma_b^{(0)}$ calculated in
the tree approximation for the free quark decay through
\be
\label{24}
\frac{d\Gamma_b}{dx}=
\frac{d\Gamma^{(0)}_b}{dx}\left(1-\frac{2\alpha_s}{3\pi}G(x)\right),
\ee
where $x=2E/m_b$. The function $G(x)$ contains the logarithmic singularities
$\sim\ln^2(1-x)$ which for $m_c=0$ appear at the quark-level endpoints
$x_{max}=1$. This singular behaviour at the end point is clearly a signal
of the inadequacy of the perturbative expansion in this region. The problem  is  solved by
taking into account the bound state effects \cite{JP94}.
Since the radiative corrections must be
convoluted with the distribution function the endpoints of the
perturbative spectra are extended from the quark level to the
hadron level and the logarithmic singularities are eliminated. In
actual calculations we neglect the terms $\sim\rho$ in $G(x)$ and
take this function from \cite{JK89}, Eq. (4.10).

The agreement with the experimental data is good. Using
$|V_{cb}|=0.042$ for the overall normalization\footnote{This value
agrees with the combined average $|V_{cb}|=(40.2\pm 1.9)\times
10^{-3}$  of Ref. \cite{CERN}.} we obtain for the semileptonic
branching ratios
\be
BR_{LF}=10.16\%,~~~BR_{ACM}=10.23\%~~~BR_{free}=10.37\%,
\ee
in agreement with the experimental finding \cite{CLEO}
$BR_{SL}=10.49\pm 0.17\pm 0.43 \%$.\\[2mm]

\noindent{\it 4. Conclusions.} We have applied a new LF formula
\cite{KNST99} to calculate the partial electron spectra in the
semileptonic $B$ decays. Using the ET and LF $b$--quark distribution
functions related by a simple kinematical transformation we compared the LF
and ACM models by computing the $b\rightarrow c$ decays for
$m_b^{LF}=<m_b^f>$. A summary of our results is presented in Table
1 and Fig.1 shows a good agreement between the results of the two
models. We have also  calculated (Fig. 2) the $b\to c$ spectrum
including the perturbative corrections and found an agreement with
the experimental data from the CLEO collaboration \cite{CLEO}. A
more detailed fit to the measured spectrum can impose constrains
on the distribution function and the mass of the charm quark. Such
the fit should also account for detector resolution.

The same formulae can be also applied for nonleptonic
$B$ decay widths (corresponding to the underlying quark decays
$b\to cq_1q_2$) thus making it possible to calculate the $B$
lifetime. A preview of this calculation can be found in \cite{KNO00}. It would be
interesting to check whether the effective values of the $b$--quark mass
can appear to be approximately the same for
different quark channels and for different
beauty hadrons. This work is in progress, and the results will be reported
elsewhere. \\[2mm]

\noindent We thank Marco Battaglia and Pepe Salt for the discussion and Karen Ter--Martirosyan
for his interest in this work. This work was supported in part RFBR grants
Refs. 00-02-16363 and 00-15-96786.

\vspace{3cm}

\noindent {\bf Table}. For the values of $p_f$ in column (1) we display
the average value of the floating b--quark mass $<m_b^f>$ (both in units of GeV) in the
second column
and the total semileptonic width of the
free b--quark (in units of ps$^{-1}$) in the third column. In the forth and fifth columns
we compare the total semileptonic widths, calculated in the ACM and LF approaches, respectively. In all cases $m_{sp}=0.15$ GeV and $m_c=1.5$ GeV and the radiative corrections are neglected. In the sixth and seventh columns, we give
the fractional deviation in percent between the semileptonic widths determined in
the LF and ACM models and that of the free quark. A momentum distribution of the b-quark is
taken in the standard Gaussian form (\ref{1}) with the Fermi momentum $p_f$. $|V_{cb}|=0.04$.\\[5mm]

{\Large
\begin{center}
\begin{tabular}{|c|c|c|c|c|c|c|}
\hline\hline $p_f$& $<m_b^f>$ & $\Gamma_{SL}^b$ & $\Gamma_{SL}^{ACM}$ &
$\Gamma_{SL}^{LF}$ & $\delta^{ACM}$ & $\delta^{LF}$ \\ \hline\hline
0.1 & 5.089 & 0.1007 & 0.1005 & 0.1005 & 0.2 & 0.2\\
0.2 & 5.004 & 0.0906 & 0.0902 & 0.0901 & 0.4 & 0.5\\
0.3 & 4.905 & 0.0799 & 0.0792 & 0.0789 & 0.9 & 1.2\\
0.4 & 4.800 & 0.0696 & 0.0688 & 0.0682 & 1.1 & 2.0\\
0.5 & 4.692 & 0.0602 & 0.0592 & 0.0584 & 1.7 & 3.0\\

\hline
\end{tabular}
\end{center}
}

\pagebreak

\begin{minipage}[t]{10.cm}
\begin{center}
\includegraphics{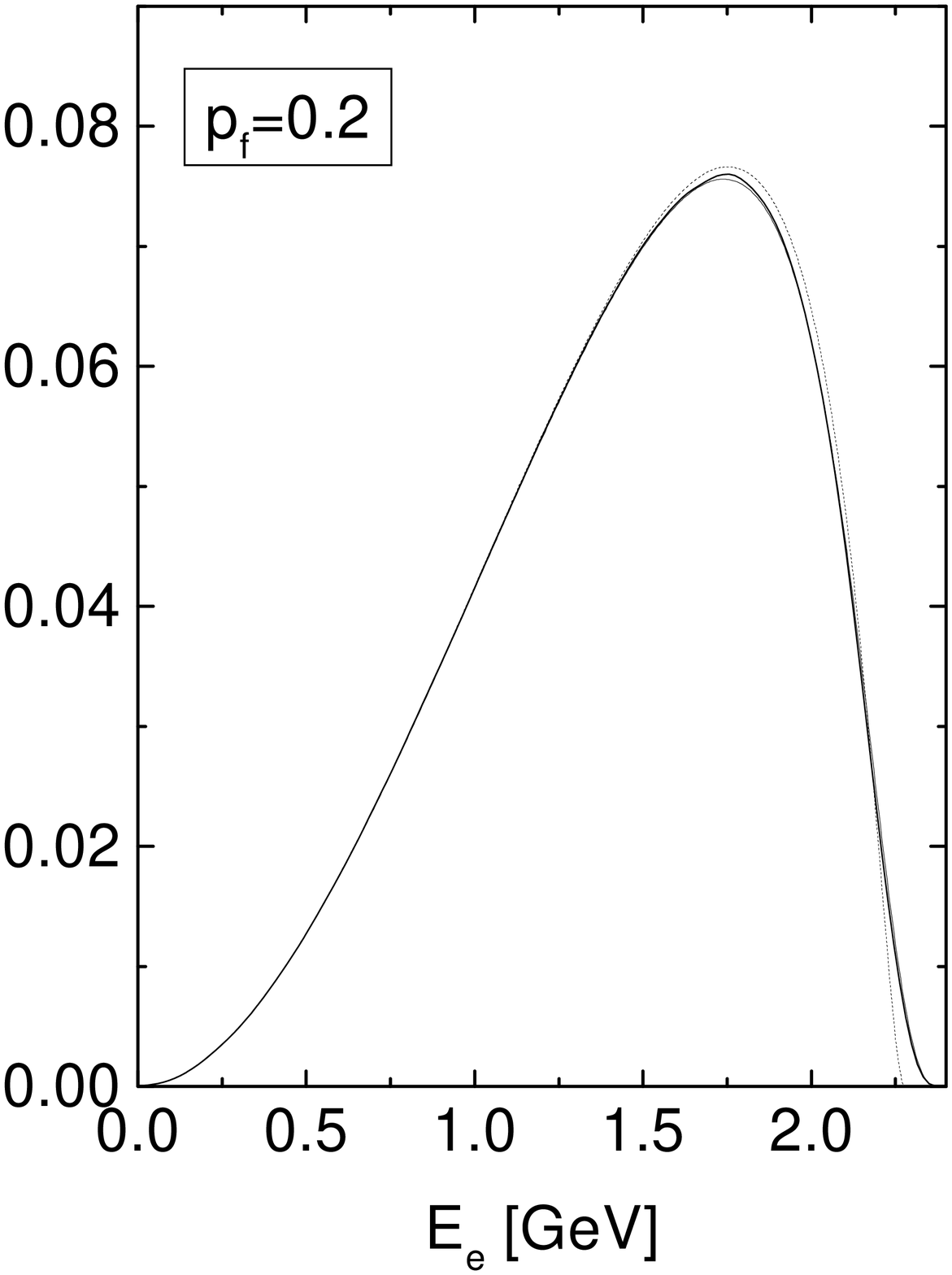}
\includegraphics{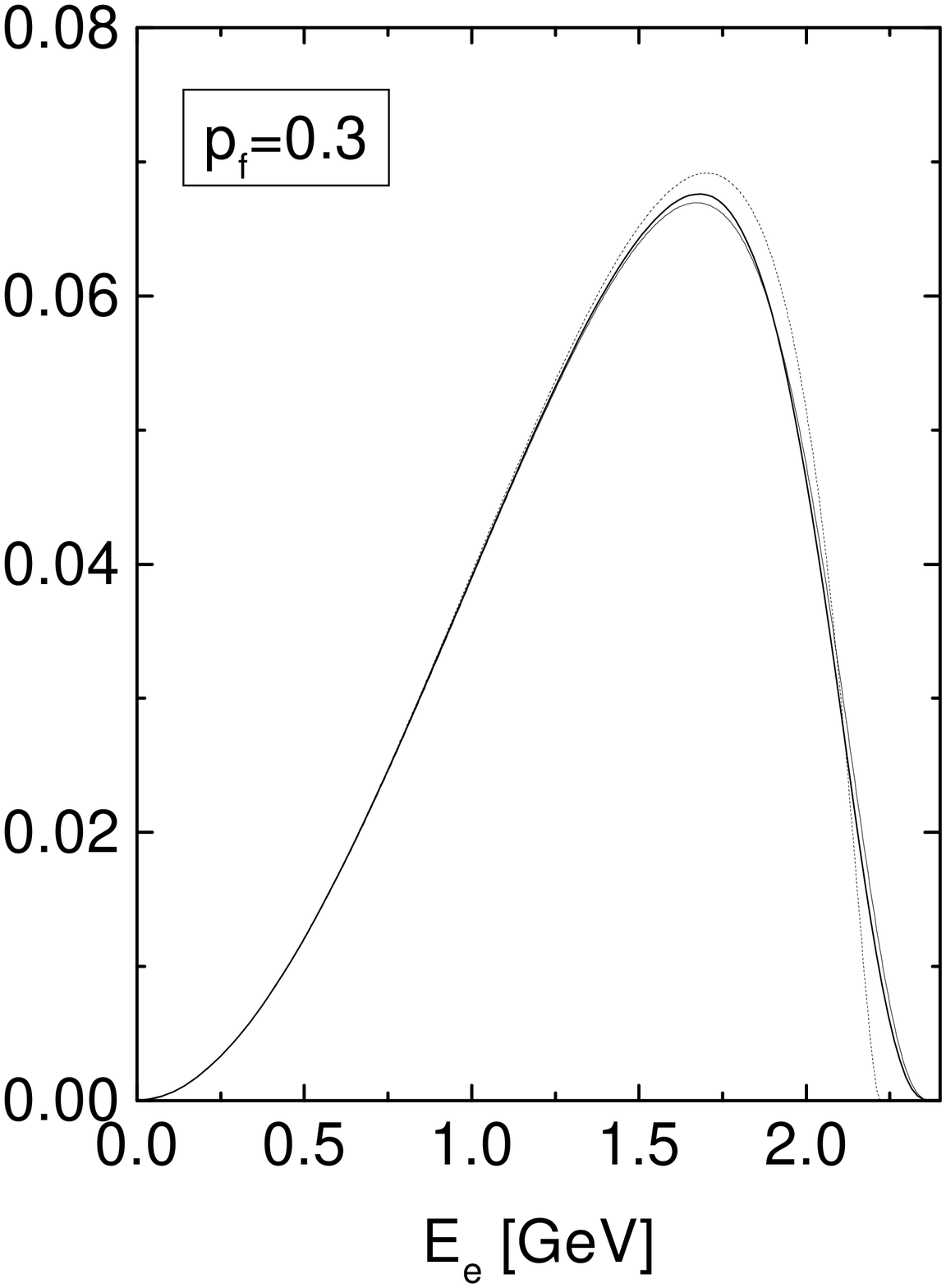}
\includegraphics{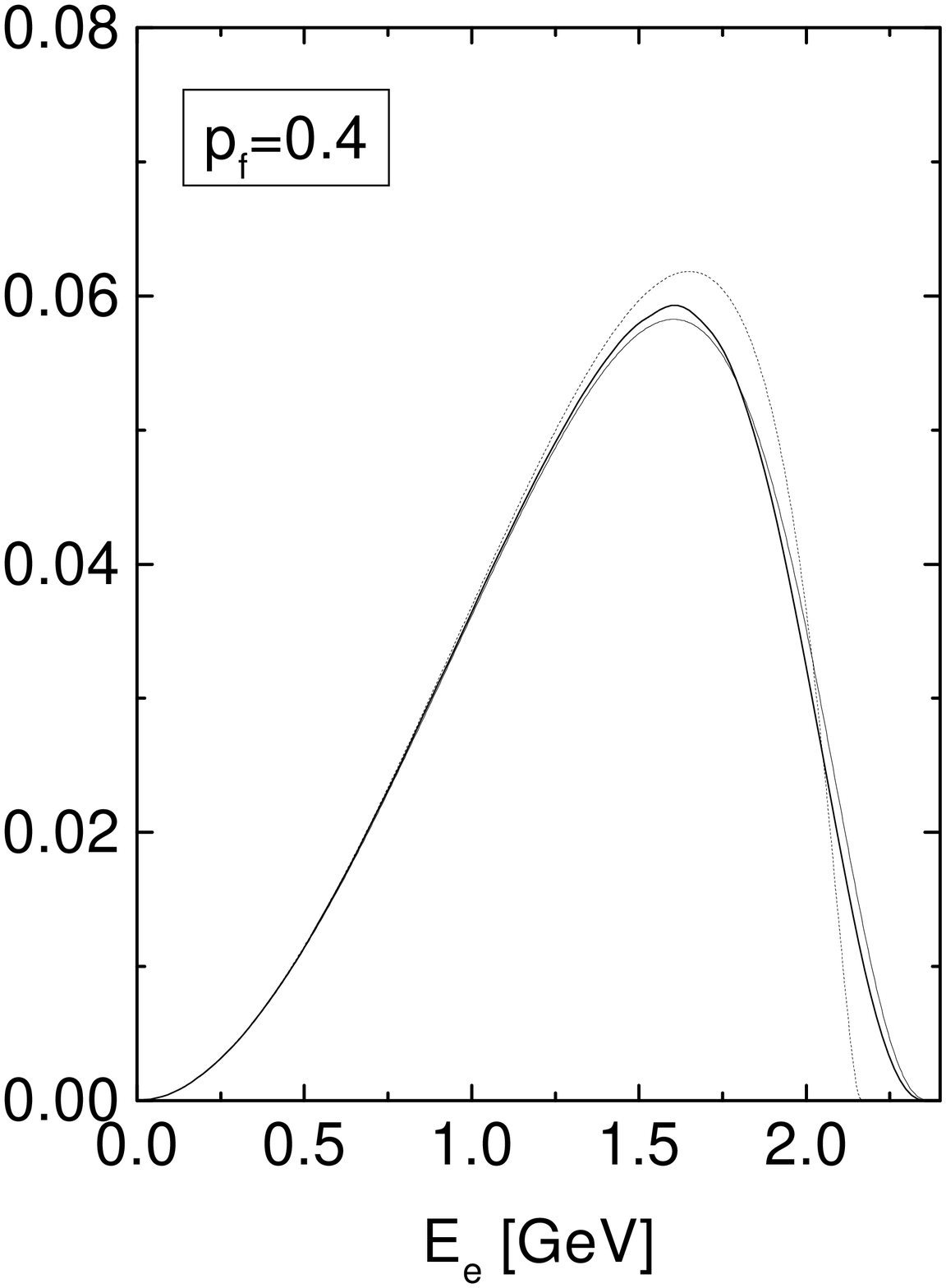}
\includegraphics{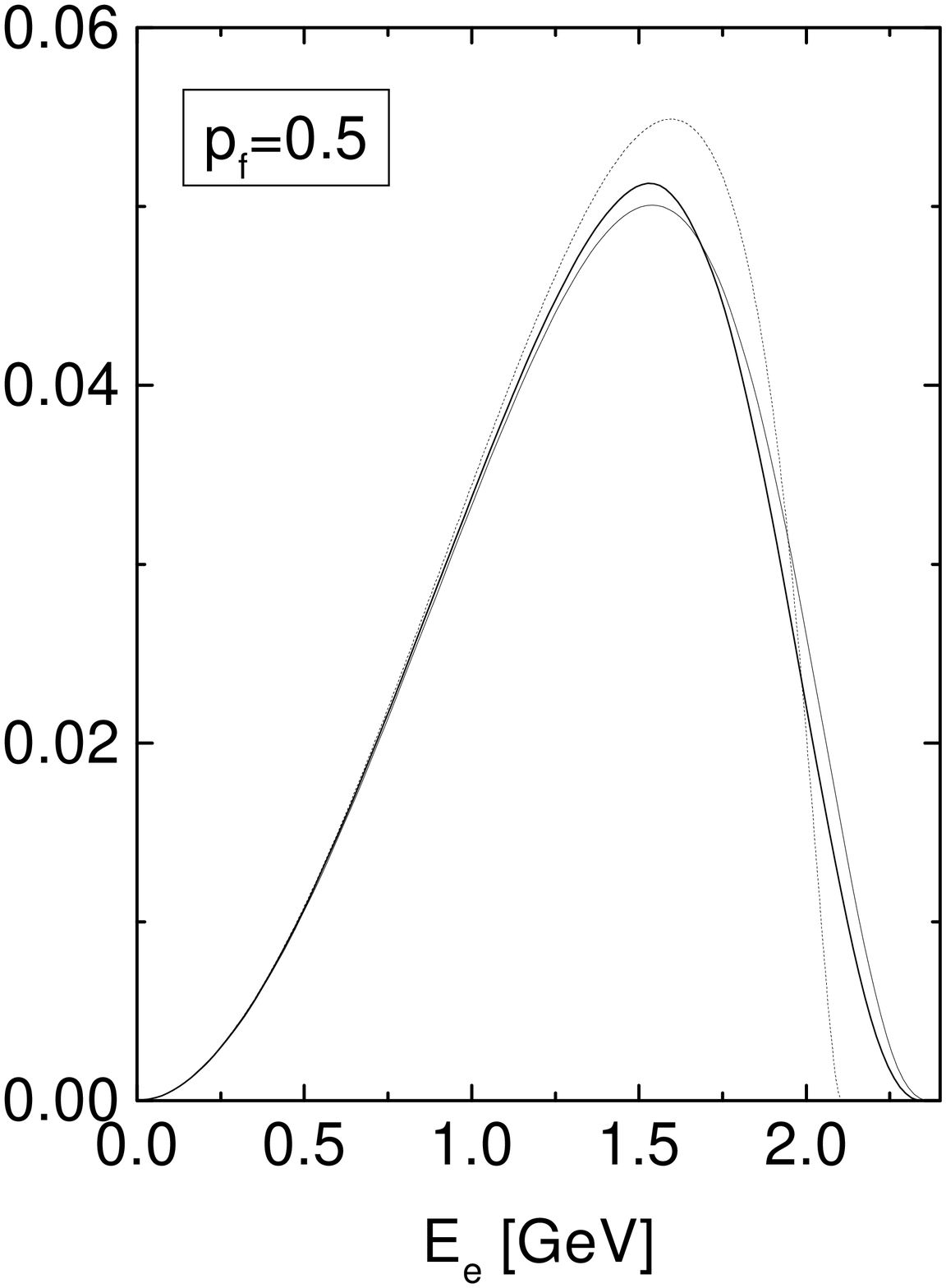}
\end{center}
\vspace*{0.8cm}
\end{minipage}
\vskip 20cm

\noindent {\bf Figure 1}. The inclusive differential semileptonic decay widths
$d\Gamma_{SL}/dE_e$ in unites ps$^{-1}$GeV$^{-1}$
for the
LF model (thick solid lines), ACM model (thin solid lines), and the free quark
decays (dashed lines). The parameters are $p_f=0.2-0.5$ GeV,
$m_{sp}=0.15$ GeV, and $m_c=1.5$~ GeV. $|V_{cb}|=0.04$.

\newpage
\begin{minipage}[t]{10.cm}
\begin{center}
\includegraphics{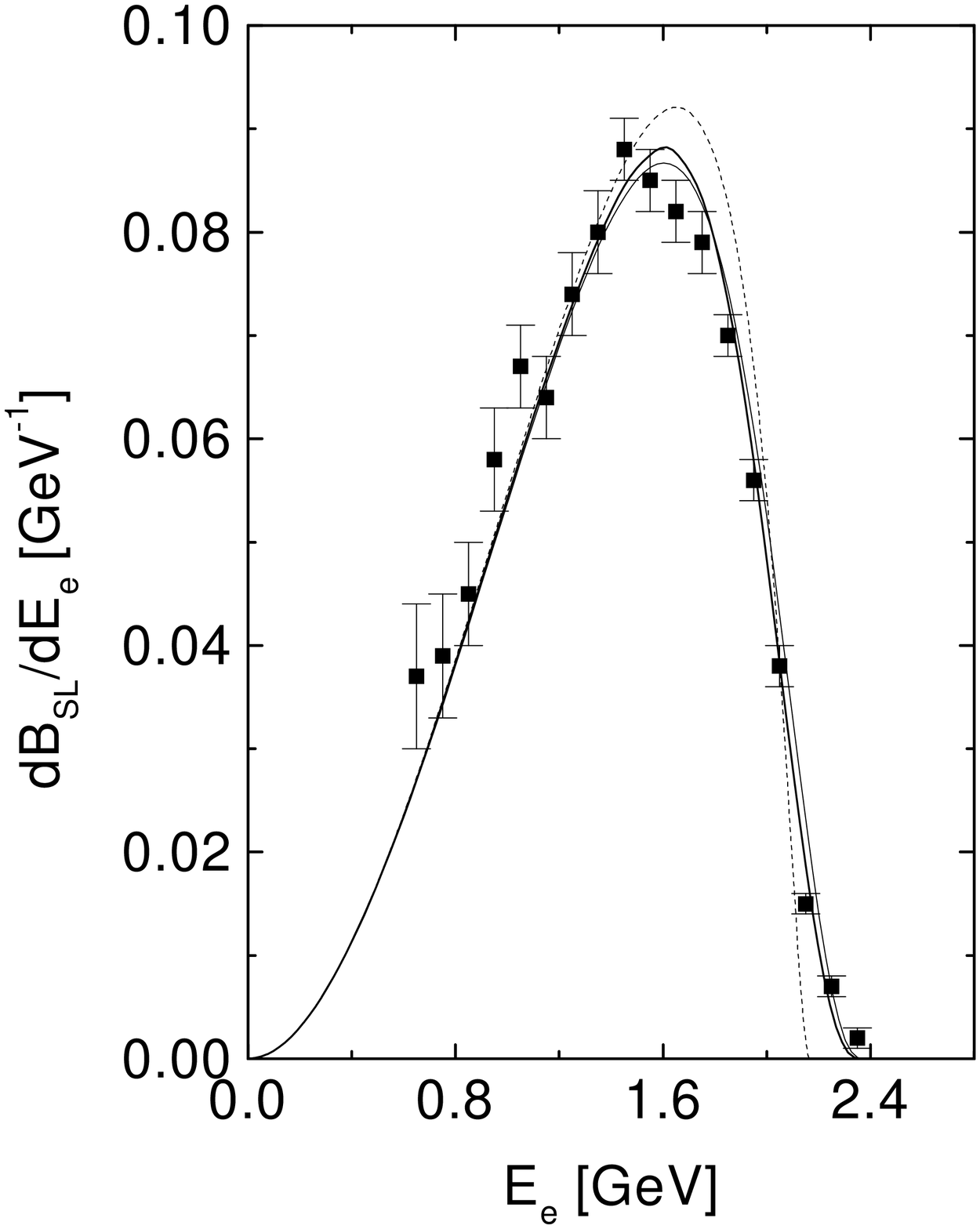}
\end{center}
\vspace*{0.8cm}
\end{minipage}
\vskip 19.8cm

\noindent {\bf Figure 2}. The predicted electron energy spectrum
compared with the CLEO data \cite{CLEO}. The calculation uses
$p_f=0.4$ GeV, $m_b=4.8$ GeV, $m_c=1.5$ GeV, and $\alpha_s=0.25$
for the perturbative corrections. Thick solid line is the LF
result, thin solid line is the ACM result, dashed line refers to
the free quark decay. The spectra normalized to $10.16\%$,
$10.23\%$, and $10.36\%$, respectively. $|V_{cb}|=0.042$.

\end{document}